\documentclass[useAMS,usenatbib]{mn2e}
\usepackage{aas_macros}
\usepackage{graphicx}
\usepackage{color}

\title[Evolution of cluster galaxies]{On the orbital and internal evolution of cluster galaxies}
\author[F.Iannuzzi and K. Dolag]{Francesca Iannuzzi$^{1}$\thanks{E-mail:
iannuzzi@mpa-garching.mpg.de} and Klaus Dolag$^{1,2}$\\
$^{1}$Max-Planck-Institut f\"{u}r Astrophysik, Karl-Schwarzschild Str.~1, D-85741 Garching, Germany\\\
$^{2}$University Observatory Munich, Scheinerstr. 1, 81679 M\"{u}nchen, Germany}
\begin{document}

\date{Accepted 2012 August 30.  Received 2012 August 16; in original form 2012 March 28}

\pagerange{\pageref{firstpage}--\pageref{lastpage}} \pubyear{2012}

\maketitle

\label{firstpage}

\begin{abstract}
Galaxies inhabiting a cluster environment experience significant evolution in their orbital motions throughout time; this is accompanied by changes in the anisotropy parameter, measuring the relative importance of radial and tangential motions for a given class of objects. Along with orbital changes, galaxies in clusters are well known to undergo severe alteration in their hot/cold  gas content and star formation properties. Understanding the link between the changes in the internal properties of galaxies and their orbital motion is of crucial importance in the study of galaxy evolution, as it could unveil the primary mechanism responsible for its environmental dependence. Do the changes in the internal properties happen in parallel with those in the orbital motion? Or are the orbital features at the time of infall what determines the fate of the member galaxies? Alternatively: are the properties of galaxies at a given time related to the coeval orbital anisotropy or are they better related to the anisotropy at infall? In order to answer these questions, we studied the orbital evolution of different galaxy populations in the semi-analytic models of Guo et al. (2011) applied on to the Millennium Simulation. For each class of objects, characterised by different internal properties (such as age, star formation rate and colour), we studied the anisotropy profile at redshift zero and its evolution by tracing the progenitors back in time. We conclude that the orbital properties at infall strongly influence the subsequent evolution of the internal features of galaxies and that the overall anisotropy of the galaxy population tends to increase with time.
\end{abstract}


\section{Introduction}
\label{sec:introduction}
In the currently favoured scenario for the formation of cosmological structure in the Universe, dark matter halos form and merge giving rise to a hierarchy of objects assembling in a bottom-up fashion. In this context, galaxy formation is pictured to arise from cooling and condensation of baryons within the potential well associated to these dark-matter structures \citep{whiterees78}. Once the galaxy is formed, its evolution will be driven by \textit{(i)} ``nature'', i.e. the object's intrinsic features (essentially stellar mass) and \textit{(ii)} ``nurture'', external processes related to the environment the galaxy inhabits during different stages of its history. It is indeed well known that, although the structural properties of galaxies are mainly determined by their stellar mass \citep{kauffmann03,tanaka04,vandenbosch08}, the existence of an environmental dependence cannot be disregarded \citep[][among the others]{hogg04,balogh04,kauffmann04,blanton05}. A combination of these effects is responsible for the observed ``bimodality'' in the galaxy properties \citep{baldry04,kauffmann04}, namely the existence of two well-distinguished classes of objects characterised by either red colour/high mass/old stellar population or blue colour/lower masses/young stellar population, the former residing preferentially in over-dense environments \citep{oemler74,dressler80,bower04,balogh04,ball08,bamford09,skibba09}, while the latter being found mainly in the field. A clear and thorough physical picture of galaxy evolution is yet to come; understanding what mechanisms play the leading role in shaping galaxy properties is a topic of ever increasing interest and research activity, but the results are still controversial. Among the environmental processes, four broad classes of mechanisms are generally considered: {\it{galaxy mergers}},  negligible in massive clusters, but important in small groups, which drive morphological changes and can affect star formation \citep{toomre72,farouki81}; {\it{strangulation}} \citep{larson80}, a sum of effects leading to the removal of the hot-gas halo associated to a galaxy when this is accreted onto a larger structure; {\it{ram-pressure stripping}} \citep{gunn72}, important in massive clusters where the density of the intra-cluster medium is highest, it leads to the progressive stripping of the gaseous component (both hot and cold) of the satellite galaxies;  {\it{tidal effects}}, arising from the gravitational interaction with other members and with the cluster potential itself, they cause stripping and heating \citep{richstone76,moore96}. Some of these processes are better understood and formalised (e.g. ram-pressure stripping), while others still lack a strict, physical description and are referred to in more generic terms (e.g. strangulation); on top of this, the regimes where each of the mechanisms is active/unimportant are only broadly assessed.  In galaxy clusters, the densest possible environments, the contribution of  mergers can be neglected, due to the large velocity dispersion of the system; strangulation occurs rapidly and devoids the hot-gas reservoir, while ram-pressure stripping and tidal interactions proceed as the satellite plunges into its host. The importance of these last two processes increases with local density (gas density for ram-pressure stripping, total matter density for tidal effects), but while in the first case this results into enhanced stripping, the effect of tidal interactions is believed to be mainly that of an induced gas consumption following the increase of nuclear activity \citep{boselli06}. Given the dependence upon gas and matter density and how these are, in turn, related to the radial distance from the cluster centre, it is obvious to expect these processes to strongly depend on the orbital history of the satellites, namely how often these happen to transit the innermost regions with respect to the outskirts. This will depend on the initial orbit of the galaxy at the time of accretion, as well as on its time evolution within the dynamically-active cluster region. There exist observational evidences that HI-deficient galaxies in nearby clusters are early-type spirals on radial orbits, while gas-rich galaxies are characterised by more tangential motions \citep{dressler86,solanes01}; other works studied the differences in the velocity distribution of early-type and late-type galaxies and found evidences of the latter moving on slightly more radial orbits, especially at large clustercentric radii \citep{mahdavi99,vandermarel00,katgert04,biviano04,biviano06,biviano08,BP09,wojtak10}. Deriving this information from a sample of observed cluster members is not an easy task and various techniques were developed appositely. By instance, knowing the projected number density profile and line-of-sight velocity dispersion of the selected galaxies, one could perform a Jeans dynamical analysis \citep{binneytremaine} to derive the cluster mass and velocity anisotropy profiles (quantifying the relative importance of radial and tangential orbits as a function of the clustercentric radius). This technique relies on the assumption of spherical symmetry, collisionless dynamics and dynamical equilibrium, in addition to being hampered by the so-called ``mass-anisotropy'' degeneracy (even if considerable progress has recently been done to remove it, see \citealt{lokas03,battaglia08, wojtak09}). This limitations, along with the difficulties in performing observations in the outskirts of clusters, should induce caution in the interpretation of the results.  Yet it would be interesting to know the extent to which the evolution of galaxies is linked to their specific orbital motion. Is the type of orbit determining the efficiency of the environmental processes in shaping the internal properties of the object? In this case we would expect severely affected galaxies to move on more markedly radial orbits, as these cross the cluster right down to its innermost regions.  Or do the changes in the galaxy properties happen in parallel to those in the orbital motion? In this case we would expect the orbits of the objects having suffered major environmental influences to differentiate from those of just-accreted satellites. \\
The importance of the numerical approach in the study of galaxy evolution need not be stressed; cosmological simulations provide a fundamental tool to understanding the assembly of structures throughout time, but the reliability regarding their treatment of interactions other than gravity is subject of debate (see, e.g., \citealt{scannapieco11}). Semi-analytic models (\citealt{white91, cole91, kauffmann93, cole94, kauffmann99, springel01a}; see \citealt{baugh06} for a review) present a powerful, hybrid approach to the problem  of galaxy formation and evolution: they make use of accurate, dark-matter-only simulations to account for the growth of structures in the cosmological context and regulate the formation and evolution of galaxies according to a set of analytical prescriptions encompassing the physics of the baryonic component. In this work we use one of the most advanced semi-analytic models developed to date to study the link between orbital and internal properties of galaxies belonging to massive clusters in a $\Lambda$CDM cosmology. We will show how the results of this analysis suggest a scenario where the specific orbital features of the satellites at the time of infall have major consequences on their evolution.
The paper is organised as follows: Sec.~\ref{sec:sams} describes the simulation, the semi-analytic model and the choice of the cluster sample used in our analysis; Sec.~\ref{sec:results}  reports the results for the anisotropy parameter of the member galaxies at redshift zero (\ref{sec:zzero}), higher redshifts (\ref{sec:highz}), at the time of the last infall inside their host (\ref{sec:infallz}) and regarding its time evolution (\ref{sec:time_evol_beta}); finally, Sections \ref{sec:summary} and \ref{sec:discussion} contain a brief summary and discussion of the results.

\section[]{The semi-analytic models and the selected sample}
\label{sec:sams}
The samples analysed in this work were extracted from the galaxy catalogue obtained by \cite{guo2011} (hereafter GUO11) after running their galaxy formation models on the Millennium Simulation \citep[hereafter MS;][]{millennium}. The MS shows how dark matter structures form and evolve in a $\Lambda$CDM cosmological scenario characterised by the parameters  $\Omega_{\rm tot} = \! 1, \; \Omega_m = \!0.25, \; \Omega_b=0.045, \; \Omega_{\Lambda}=0.75, \; h = 0.73, \; \sigma_8=0.9, \; n_s=1$. The evolution of cosmological structures is traced by $2160^3$ particles moving in a periodic box of side $500$ Mpc/h under the mutual gravitational influence. The results of the simulation were stored at $64$ different times starting from redshift $127$ down to redshift zero. At each of these times a catalogue of bound structures and substructures was generated, by applying a friend-of-friend technique \citep{fof} and the SUBFIND algorithm \citep{springel01a}  on the particle distribution; these catalogues constitute the basis for recovering the merger history of structures throughout time, also referred to as the merger tree of the simulation. Semi-analytic models of galaxy formation, as those of GUO11, provide a description of the cosmological evolution of baryonic matter; a gas distribution is associated to each bound structure identified in the simulation and its evolution in time is regulated by a set of recipes implemented on to the dark-matter merger tree identified in the simulation: according to the specific histories of each of the substructures, the evolution of the associated baryonic components will follow its own, peculiar path. The physical processes implemented in the recipes of GUO11 include cooling, star formation, supernova and AGN feedback, hot-gas stripping, metal enrichment and alone they provide a remarkable match to the abundance and large-scale clustering properties of the observed galaxy population at low redshift. We will comment on the limits of the model in the discussion of our results; we refer the reader to GUO11 for a thorough description of the implemented physics and of the strengths and weaknesses of their approach.\\
We have selected the $1000$ most massive clusters identified in the MS at redshift zero; these have virial masses\footnote{These are defined as the masses of the spherical regions centred on the potential minimum of the smooth halo and corresponding to an overdensity of $\Delta$, typically $\approx 200$, with respect to either the critical density $\rho_{crit}$ or to the mean background density $\rho_{m} = \Omega_m\rho_{crit}$. Hereafter, when referring to either the virial radius or the virial mass of an object, we assume the overdensity to be defined with respect to the critical density.} greater than $2 \times 10^{14} \; M_{\odot}$ and contain a total of around one million galaxies. Although all of these galaxies are originally associated to a specific dark-matter subhalo, not all of them preserve their dark-matter component throughout the simulation time; this means, in turns, that in these cases the dynamical evolution can no longer be provided by the underlying simulation. The fate of these \textit{orphan} galaxies (generally referred to as ``type $2$s'' in the galaxy formation model), which have lost their dark-matter component, is to eventually merge with the galaxy at the centre of the cluster they inhabit, within the timescales set out by dynamical friction. During this time, the orbit of the orphan galaxy is traced by the most bound particle present in its dark matter halo before this vanished, modified by a shrinking factor introduced to mimic the orbital decay caused by dynamical friction. Since the orbits of these galaxies are altered, we decided not to include them in our analysis. This leaves us with $\approx 2\times10^5 $ galaxies in the mass range $10^3 < M_*/M_{\odot} < 10^{12}$, where $M_*$ indicates stellar mass.
We have stacked the resulting galaxy sample by subtracting bulk motions and by normalising positions and velocities to the virial values. To summarise, we are studying the mean behaviour of galaxies residing in the most massive clusters at redshift zero.\\
From the galaxy catalogue we have access to a wealth of information regarding the internal properties of the selected objects; those we are primarily interested in are stellar/dark matter masses and colour, the latter defined as the the difference between the rest-frame total absolute magnitude in the SDSS $u$ and $i$ bands. We will eventually discuss the impact on our results of using other quantities, such as mean stellar age and specific star formation rate, to split our galaxy sample into distinct populations.

\section[]{Results}
\label{sec:results}
In this section we will present the results on the velocity anisotropy for the galaxies in the selected sample. This quantity is defined as:
\begin{equation}
\beta = 1-\frac{\sigma^2_t}{\sigma^2_r},
\end{equation}
where $\sigma_t$ and $\sigma_r$ are the velocity dispersions in the tangential and radial direction\footnote{By tangential direction we mean that identified by variations in any of the angles $\theta$ or $\phi$, i.e. $\sigma_t^2 = (\sigma_{\theta}^2+\sigma_{\phi}^2)/2$.}, respectively. The velocity anisotropy can take values ranging from $-\infty$ to $1$; the former case corresponds to purely circular motions (no dispersion in the radial direction), while in the latter the orbits are completely stretched along the radial direction (no tangential dispersion). The case $ \beta \approx 0 $ is referred to as {\it{isotropy}} and it occurs when the velocity dispersion is of comparable magnitude in both radial and tangential directions. We will mostly discuss global values of $\beta$, describing a full population of objects without discriminating the spatial distribution of the members; in one case we will also show the radial profile, to underline both the general properties of the anisotropy parameter and its dependence on the distance form the centre of the cluster. In what follows, we split the galaxy population into a red and blue sample according to the value of the $u-i$ colour indicator, namely whether it is greater or smaller than a certain threshold; this is set to the value at which the colour distribution of the galaxy sample splits into two well-defined, different components. At redshift zero, this happens around $u-i=2.5$; considering all the galaxies within three virial radii, the resulting ``red'' sample consists of $\approx 72000$ objects, against the $\approx 130000$ of the ``blue'' counterpart\footnote{The higher fraction of blue objects can be traced back to \textit{(i)} the exclusion of type $2$s and \textit{(ii)} the choice of three virial radii. When restricting the analysis to $1.5$ virial radii and including type $2$s, the number of red objects becomes nearly four times as large as that of the blue ones. }.

\subsection{Anisotropy at redshift zero}
\label{sec:zzero}
\begin{figure}
\includegraphics[width=84mm, angle=0]{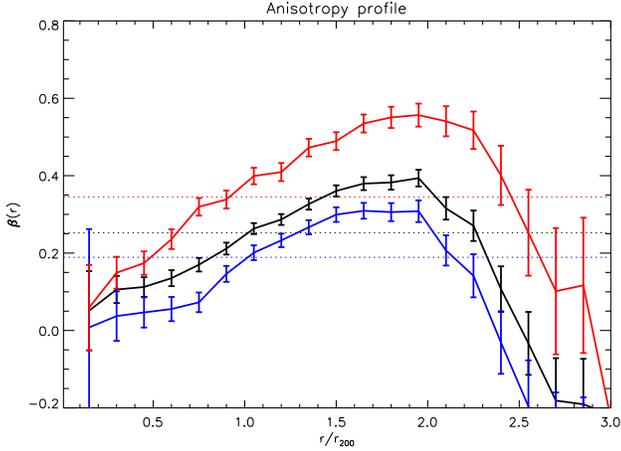}
 \caption{Anisotropy profile for the stacked sample at redshift zero. The black, solid  curve shows the radial behaviour of the anisotropy parameter for the full population of galaxies found in the selected cluster sample; the red (blue), solid curve represents the result for the subsample of galaxies characterised by an $u-i$ colour greater (smaller) than $2.5$. The error bars are evaluated via a bootstrapping algorithm and correspond to two standard deviations. The horizontal, dotted lines mark the global value of the anisotropy parameter for all the galaxies within three virial radii.}
 \label{beta_profile}
\end{figure}

We first analyse the results at redshift zero and as a function of the radial distance of the galaxies from the centre of the stacked cluster. The curves in Fig.~\ref{beta_profile} show the behaviour of the anisotropy parameter in $20$ equally-spaced radial shells, extending from roughly $0.1$ to $3$ virial radii. The error bars were calculated by means of a bootstrapping algorithm and correspond to two standard deviations (this holds for all the uncertainties reported in this paper, unless explicitly stated otherwise). The black curve shows the result for the full population: the importance of radial motions increases, moving from the central regions outwards and peaks around $1.5-2$ virial radii. This behaviour is typical of systems formed by gravitational collapse \citep{vanalbada82} and it is compatible with other results in the literature \citep{rasia04,gill04,mamonlokas05,sales07,wojtak09,BP09,host09,lemze11,lapi11}. The blue and red curves represent the results for the subgroups of galaxies characterised by blue and red colours, respectively; the clear message coming from the plot is that the blue population has a systematically and significantly lower anisotropy than the red population, at least between  $0.5$ and $2.5$ virial radii. This is confirmed by the global values of $\beta$, marked by the horizontal, dotted lines. These were computed out of all the galaxies found at distances less than three virial radii from the centre, with the following result: $ \beta_{all} = 0.253 \pm 0.006 \;, \beta_{blue}=0.189 \pm 0.006\;, \beta_{red}=0.345 \pm 0.008$. In summary, red galaxies in our sample move on more radially-biased orbits than the blue counterparts. A clear explanation of this effect will progressively arise in the following sections.
\subsection{Anisotropy at high redshift}
\label{sec:highz}
\begin{figure}
\includegraphics[width=84mm, angle=0]{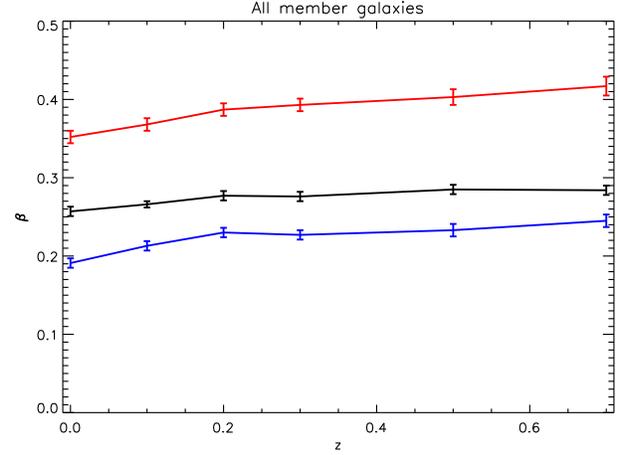}
 \caption{Global value of the anisotropy parameter as a function of redshift. Considered are all the member galaxies belonging to the high-redshift progenitors of the redshift-zero cluster sample; these may include objects which do not survive to redshift zero. The black curve shows the result for the full galaxy population at each redshift, whereas the red (blue) curve refer to the subsample of objects with $u-i$ colours greater (smaller) than a threshold; this is set to be the value at which the colour distribution splits into two, well-defined component (e.g. $u-i \approx 2.5$ at redshift zero, $u-i\approx2.35$ at $z=0.7$). The error bars are evaluated via a bootstrapping algorithm and correspond to two standard deviations.}
 \label{beta_highz_all}
\end{figure}

\begin{figure}
\includegraphics[width=84mm, angle=0]{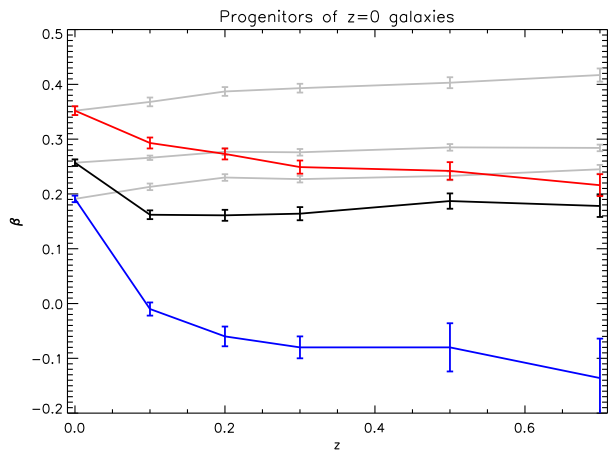}
 \caption{Global value of the anisotropy parameter as a function of redshift. Considered are the progenitors of the galaxies belonging to the redshift-zero cluster sample; only progenitors which are already satellites and will remain satellites down to redshift zero are taken into consideration. The black curve shows the result for the full population at each redshift, whereas the red (blue) curve refer to the subsample of objects with redshift-zero descendant characterised by $u-i$ colours greater (smaller) than $2.5$. Overplotted in gray are the results displayed in Fig.~\ref{beta_highz_all}. The error bars are evaluated via a bootstrapping algorithm and correspond to two standard deviations.}
 \label{beta_highz_prog}
\end{figure}

In what follows, we focus our attention on the time evolution of the anisotropy parameter. Having not found significant changes in the radial behaviour with respect to the trend shown in Fig.~\ref{beta_profile}, we will only refer to global values of $\beta$ hereafter. We proceeded in two ways: we considered all the member galaxies belonging to the high-redshift progenitors of the selected clusters and, in parallel, we also analysed the progenitors of the redshift-zero galaxies only. The two approaches differ in that the first galaxy sample may contain objects that do not survive until redshift zero, whereas in the second case only satellites that have a redshift-zero descendant are considered. We show the results of the first method in Fig.~\ref{beta_highz_all}, where the value of $\beta$, computed for the full population of member galaxies regardless of their spatial position, is plotted as a function of redshift. The colour code is the same as for Fig.~\ref{beta_profile}; keeping the threshold $u-i=2.5$ for the splitting between the red and blue population or adjusting it to the variation in the colour distribution of the high-redshift galaxies introduces no substantial changes in the results. \\
The two main considerations arising from this analysis are \textit{(i)} that the global value of the anisotropy parameter does not significantly evolve in time and \textit{(ii)} that the blue population is characterised by a lower degree of anisotropy at all redshifts. 
A more interesting result is given by the analysis of the second sample of high-redshift galaxies. These are progenitors of redshift-zero objects that are already members of their final, host cluster at the redshift of interest and that will not leave it anytime afterwards. Fig.~\ref{beta_highz_prog} shows the global anisotropy for this class of galaxies; the colour code refers to the redshift-zero population only and not to the progenitors: the red (blue) curve corresponds to progenitors of galaxies that are red (blue) at redshift zero. The plot shows with striking clarity that, at each redshift, the anisotropy of the progenitors of redshift-zero galaxies is considerably lower than the value for the whole, high-redshift population (Fig.~\ref{beta_highz_all}, overplotted in gray); this effect seems to be even stronger for progenitors of galaxies that are blue at redshift zero. These results suggest \textit{(i)} that galaxies on more radial orbits have less chances to survive and are progressively removed from the member population and \textit{(ii)} that a similar selection effect acts on the galaxy colour (namely: galaxies on radial orbits have very low chances to remain blue until redshift zero). The second point will become even clearer when analysing the anisotropy at the time the galaxies fall inside their host cluster.
\subsection{Anisotropy at infall}
\label{sec:infallz}
\begin{figure}
\includegraphics[width=84mm, angle=0]{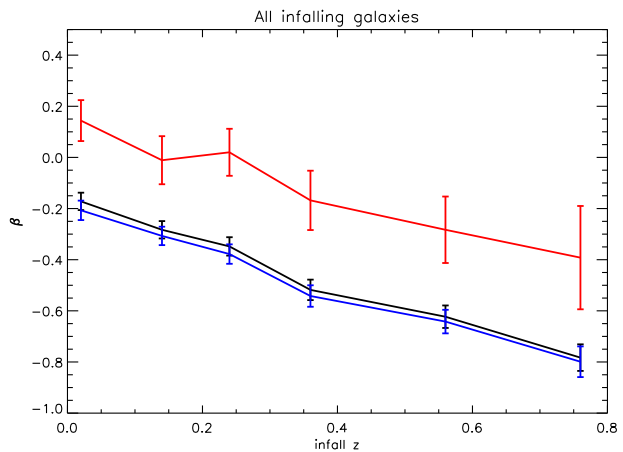}
 \caption{Global value of the anisotropy parameter for the infalling population, as a function of infall redshift. Considered are all the galaxies falling inside the progenitors of the redshift-zero clusters at different times; these may include objects which do not survive to redshift zero. The black curve shows the result for the full infalling population at each redshift, whereas the red (blue) curve refer to the subsample of objects with $u-i$ colours greater (smaller) than $2.5$. The error bars are evaluated via a bootstrapping algorithm and correspond to two standard deviations. }
 \label{beta_infall_all}
\end{figure}

\begin{figure}
\includegraphics[width=84mm, angle=0]{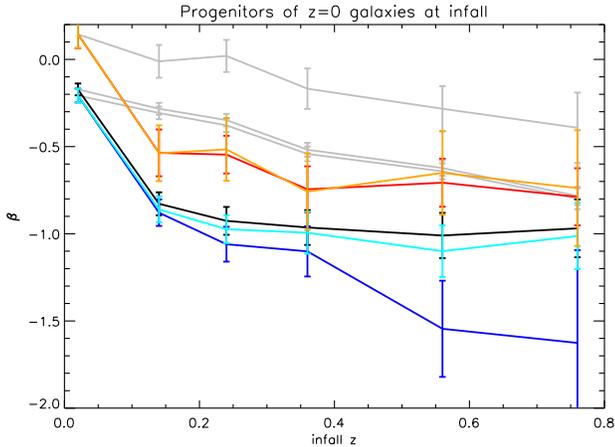}
 \caption{Global value of the anisotropy parameter for the infalling population, as a function of infall redshift. Considered are the progenitors of the galaxies belonging to the redshift-zero cluster sample; only progenitors which are becoming satellites for the last time in their history are taken into consideration. The black curve shows the result for the full infalling population at each redshift, whereas the red (blue) curve refer to the subsample of objects with redshift-zero descendant characterised by $u-i$ colours greater (smaller) than $2.5$. Displayed in orange (cyan) is the anisotropy of the subgroups of infalling satellites which are red (blue) at the time of infall. Overplotted in gray are the results displayed in Fig.~\ref{beta_infall_all}. The error bars are evaluated via a bootstrapping algorithm and correspond to two standard deviations.}
 \label{beta_infall_prog}
\end{figure}

We finally study the orbital properties of galaxies at the time they become members of their cluster; by this we mean the moment the dark-matter halo of a galaxy is not an independent structure anymore, but a substructure identified within a larger halo. We want to see how the anisotropy of the infalling population varies with time. Again, we split the analysis into two parts; first, we consider the full sample of galaxies joining the high-redshift progenitors of our original cluster sample and, second, we also restrict the study to the subsample of galaxies with a descendant at redshift zero. The results of the first approach are displayed in Fig.~\ref{beta_infall_all}, where we plot $\beta$ against the infall redshift. The red (blue) curve refers again to galaxies that are red (blue) at the redshift under consideration; keeping $u-i = 2.5$, or varying the $u-i$ colour threshold to adapt to the colour evolution of the galaxy population at high-redshift introduces non substantial changes in the results. A clear trend appears, showing that the infall anisotropy increases going towards lower redshifts; this is not surprising, as mass-accretion is expected to occur progressively along small filaments, extending radially outside massive clusters. The second feature which is apparent in the plot is that most of the infalling population consists of blue galaxies, as shown by the vicinity of the black and blue curve, as well as by the size of the error bars. Third we see that, again, the anisotropy of blue galaxies is lower than that of the red galaxies. At a first glance, this may result somewhat unexpected as there is no obvious reason why such a trend should already be in place before entering the cluster environment. We note, though, that the mean mass of the red sample is considerably higher than for the blue counterpart (almost two order of magnitudes for both stellar and dark-matter masses); previous works have already stated that more massive satellites tend to approach the cluster along more eccentric orbits, as they are more likely to reside in dense, radial filaments than the least massive ones \citep{tormen97}. This effect could explain the behaviour observed in Fig.~\ref{beta_infall_all} and also play a role in the interpretation of Fig.~\ref{beta_profile}; an initially higher anisotropy for the red population of infalling galaxies may leave an imprint in the final profile at redshift zero. The impact of this effect is anyway limited by the small number of these galaxies within the infalling population: at $z=0.7$, out of $19049$ infalling galaxies only $419$ are red. Clearly, these objects may only represent a contribution to the population of red, member galaxies at low redshifts.
Fig.~\ref{beta_infall_prog} shows the results for the subsample of infalling galaxies with a redshift-zero descendant. As usual, in black are the result for the full population whereas the red (blue) curve refers to the subsample of galaxies that are progenitors of objects characterised by a red (blue) colour at redshift zero. Again we see that, at each redshift, the anisotropy of these galaxies is systematically lower than the value for the full, infalling population (Fig.~\ref{beta_infall_all}, overplotted in gray). This suggests that galaxies present in the cluster at redshift zero tend to originate from the subsample of objects entering the cluster environment with the least radially stretched orbit. We then split the full population not according to the colour of the redshift-zero descendant, but on the basis of their colour at the time of infall; the orange (cyan) curve represent the infall anisotropy of galaxies that are red (blue) at the time of infall. We clearly see that the galaxies which are blue at redshift zero are descendant of the subgroup of infalling, blue galaxies with the lowest infall anisotropy. It seems, therefore, that not only does the initial orbit strongly influence how long the galaxy is going to survive in a cluster environment, but that it also plays an important role in determining the evolution of its internal properties. 

\subsection{Evolution of anisotropy in time}
\label{sec:time_evol_beta}
\begin{figure}
\includegraphics[width=84mm, angle=0]{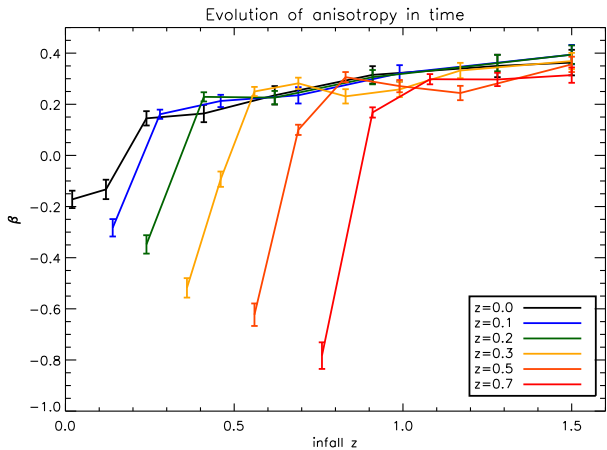}
 \caption{Global value of the anisotropy parameter at different times and as a function of infall redshift. Each curve represents the results at a specific redshift, as stated in the legend. The points on the curves correspond to the anisotropy of galaxies which have last become satellites at the redshift given on the x axis. The error bars are evaluated via a bootstrapping algorithm and correspond to two standard deviations.}
 \label{beta_time}
\end{figure}

\begin{figure}
\includegraphics[width=84mm, angle=0]{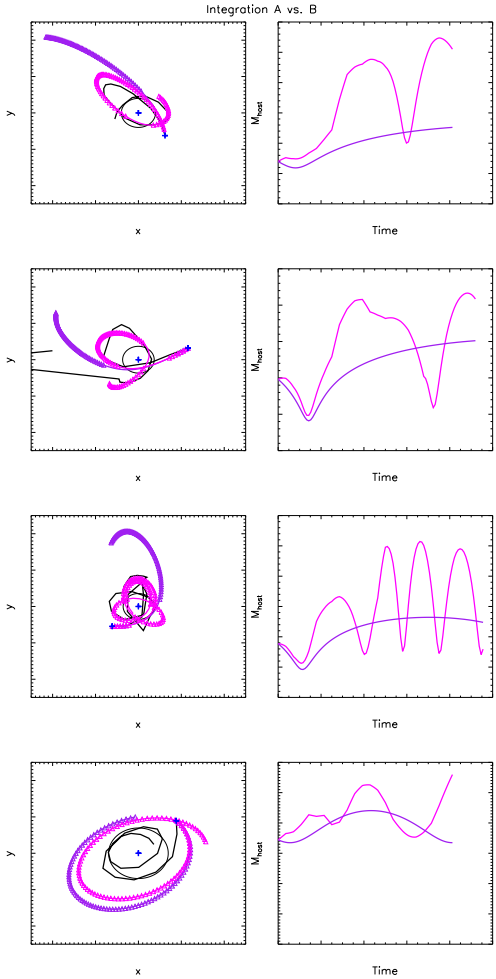}
 \caption{Examples of integrated orbits. The left column shows the motion of four different galaxies around their host. The blue crosses represent the centre of the host and the initial position of the galaxy, while the circle gives the size of the virial region at the beginning of the integration. The black curve corresponds to the original orbit from the MS and the purple (magenta) curve shows the result of the integration in case A (B). Whenever the orbit is bound, a triangle is overplotted; this is not done for the original orbit, as the information on its energy is not available. The timestep is $90$ Myrs, approximately one fourth of the temporal separation between two MS snapshots ($\approx 350$ Myrs). The right column shows the evolution of the mass felt by the galaxy as it orbits around the host in both case A (purple curve) and case B (magenta curve); as spherical symmetry is assumed in the mass distribution of the host, this quantity corresponds to the mass contained within the sphere of radius $\bmath{r} = \bmath{r}_{sat} - \bmath{r}_{host}$.}
 \label{orbits}
\end{figure}

\begin{figure}
\includegraphics[width=84mm, angle=0]{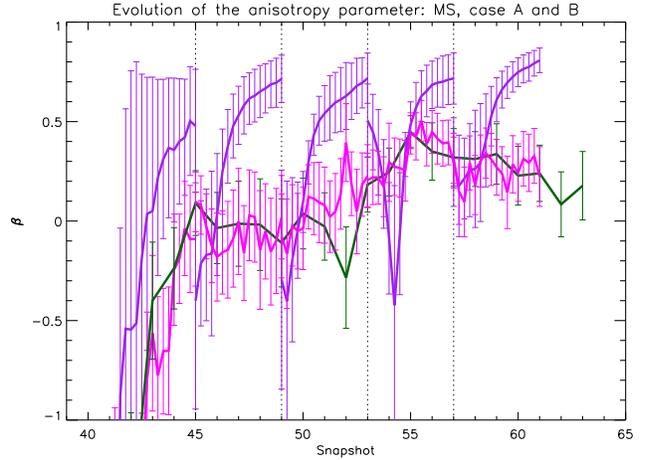}
 \caption{Evolution of the anisotropy parameter as obtained from the MS (green curve) and for cases A (purple curve) and B (magenta curve). The integration of the orbits has been performed on a randomly selected subsample of approximately $100$ galaxies sharing the same infall snapshot, given their initial position and velocity. The results of the integration are compared to the profile computed out of simulated data for the same galaxies. The plot shows the temporal evolution of $\beta$ for the whole population, as a function of snapshot number -- from $41$ (i.e. $z=1$, the infall redshift) to $63$ (i.e. $z=0$). The integration has been split into five slots extending roughly $1$ Gyr in time; at the beginning of each slot (marked by a vertical, dotted line) the position and velocity of the satellite was updated to the real value obtained from the MS. The error bars are evaluated via a bootstrapping algorithm and correspond to one standard deviation.}
 \label{integration_beta}
\end{figure}

\begin{figure}
\includegraphics[width=84mm, angle=0]{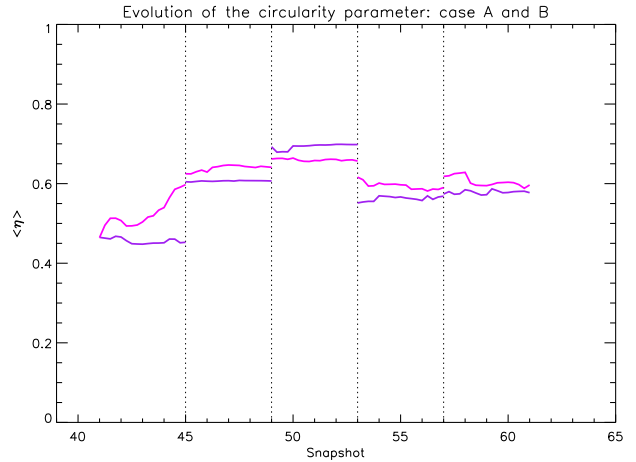}
 \caption{Evolution of the circularity parameter for the subsample of bound orbits obtained by integrations A (purple curve) and B (magenta curve). The integration of the orbits has been performed on a randomly selected subsample of approximately $100$ galaxies sharing the same infall snapshot, given their initial position and velocity. The plot shows the temporal evolution of the mean of the circularity distribution for the bound population, as a function of snapshot number -- from $41$ (i.e. $z=1$, the infall redshift) to $63$ (i.e. $z=0$). The integration has been split into five slots extending roughly $1$ Gyr in time; at the beginning of each slot (marked by a vertical, dotted line) the position and velocity of the satellite was updated to the real value obtained from the MS.}
 \label{integration_eta}
\end{figure}

When visually comparing the results for the infall anisotropy (Fig.~\ref{beta_infall_all} and \ref{beta_infall_prog}) to the typical values at redshift zero (Fig.~\ref{beta_profile} and first points of Fig.~\ref{beta_highz_all}) we already guess that $\beta$ increases, from the time the galaxies enter the cluster to the end of the simulation. This is shown more quantitatively in Fig.~\ref{beta_time}. Each of the curves represents the results at a specific redshift, from $0.7$ (in red) to zero (in black); the points on each curve correspond to the anisotropy, at this redshift, of the subgroup of galaxies that fell inside the clusters at the redshift given in abscissa. The first points of the lines show the anisotropy of the galaxies which are just infalling and, taken altogether, they reproduce the black curve of Fig.~\ref{beta_infall_all}. Overall, there is a strong indication of an increase of $\beta$ with time, after the galaxies become satellites and orbit within the environment of a larger halo; this seems to occur rapidly within the first $2$ Gyrs after infall and more gradually afterwards.\\
In order to understand what this increase was due to, we investigated the evolution of individual orbits. From a random selection of $500$ galaxies belonging to the redshift-zero sample, we isolated $\approx 100$ objects sharing the same infall snapshot and have recovered their full orbital history. Knowing the satellites' mass, their position, velocity and the virial mass of the hosts at the time of infall, we integrated the orbits forward in time. We used the leapfrog method to solve for the motion of the objects, with timesteps of around $90 \;\textrm{Myrs}$. This time resolution corresponds, approximately, to one fourth of the temporal separation between the different snapshots in the MS (reducing the size of the timestep further does not significantly change the results). The integration proceeds for a number of steps corresponding to the total time elapsing from infall ($z=1$) to redshift zero. We adopt two different approaches in the integration: in the first case (hereafter case A) we keep the mass of both the host and the satellite constant at the infall value, whereas in the second case (hereafter case B) we update both masses according to the values provided in each of the snapshots\footnote{The value for the dark-matter mass of the satellite, recovered from the simulation, is subject to inaccuracies whose importance depends on the spatial position of the object (see \citealt{knebe11}). We have performed the integration updating the mass of the host only and keeping the mass of the satellite fixed at the infall value: the differences with the results obtained in case B are negligible.} and by means of a linear interpolation to account for our finer temporal integration. In both cases we assume the mass of the host to follow a spherical NFW profile \citep*{nfw}. The concentration parameters were evaluated from Eq.~5 of \cite{neto07}, who obtained the concentration-mass relation for the halos identified in the MS at redshift zero; we do also allow the concentration parameter to vary with redshift and assume the dependency found by \cite{duffy08} for simulated halos in the redshift range $0-2$. In both integrations the satellites are regarded as point masses and no treatment of dynamical friction is included; this should not significantly impact the shape of the orbits, though \citep{vdb99,penarrubia05}. At each timestep, the mass felt by the orbiting satellites varies according to the distance from the centre of the host (for both case A and B) and to the cosmological accretion (taken into account in case B only). Case A hardly ever reproduces an orbit close to the original, whereas case B provides a better description to the motions occurring in the simulation. Even in this case, however, a perfect match to the original orbit is not guaranteed. Our approximation for the distribution of the host mass, on top of the fact that we are neglecting the effect of interactions with other orbiting galaxies, can sometimes result in evident mismatches\footnote{For a thorough account of the difficulties in reproducing satellite orbits, see, e.g., \cite*{lux10}.}. Fig.~\ref{orbits} shows few examples of original orbits from the MS (black curves in the plots on the left column) and the results of our integration for case A and B (overplotted purple and magenta curves, respectively); as mentioned, in some cases our approximation grasps the original dynamics pretty well (e.g. magenta curves in the first three rows), while in others we are clearly missing something (e.g. last row). The right column shows the time evolution of the mass felt by the satellite in both case A and B and explains the differences in the corresponding orbital evolution. \\ 
Notwithstanding the simplicity of our model, we can reproduce fairly well the evolution of the anisotropy parameter for the selected galaxies. This is shown in Fig.~\ref{integration_beta}, where $\beta$ is plotted against the snapshot number, the latter ranging from $41$ (i.e. $z=1$, the infall redshift) to $63$ (i.e. $z=0$, the end of the simulation). The green curve shows the result obtained from the simulation catalogue, i.e. the real temporal evolution of the anisotropy parameter for the selected galaxy sample. In purple (magenta) are the results from case A (B). We have split the integration into five slots lasting each around $1\; \textrm{Gyrs}$; at the end of each slot, we updated the position and velocity of the satellites from the real value obtained from the MS.
In case A, $\beta$ immediately increases towards one after the integration is switched on; this is due to most of the galaxies becoming unbound and can be related to a progressive underestimation of the depth of the potential well they are moving in. Case B, instead, provides a remarkable match to the original profile. The agreement is maintained to a reasonably good level also when the integration is let proceed uninterrupted from infall to redshift zero.\\
Interestingly, we found that this trend does not translate naively into the corresponding behaviour of the circularity distribution. Knowing position, velocity and masses one can straightforwardly compute the circularity parameter:
\begin{equation}
\eta = \frac{L}{L_c},
\end{equation}
where $L$ is the angular momentum of the orbit and $L_c$ is that of the circular orbit characterised by the same orbital energy $E$. The circularity parameter has finite values only for bound orbits, i.e. those with $E<0$; the limit $\eta = 0$ corresponds to purely radial orbits, whereas $\eta = 1$ to perfectly circular motions. At each timestep of the integration, we computed the circularity parameter for each of the bound orbits and looked at its evolution with time. Fig.~\ref{integration_eta} shows the temporal evolution of the mean of the circularity distribution, for both integration A and B. As in Fig.~\ref{integration_beta}, the results are shown for five integration slots, at the beginning of which the position and velocity of the satellite is updated to the simulation value. The purple curves, corresponding to case A, are flat: the mean circularity does not evolve with time. The same holds for the magenta curves, corresponding to case B, which also show no net evolution, other than for a mild increase in the first slot. The slope is what is significant here, while the normalisation of the curves and the differences among them are not strictly meaningful. Indeed, the first point of each slot is computed out of position and velocities extracted from the simulation, while the potential is provided by our simplified model; this makes the overall evaluation of the circularities not fully consistent. When letting the integration run uninterrupted from infall to redshift zero, the results are again consistent with no net evolution of the mean circularity of the galaxy population for both integrations. This analysis is performed on the bound subset of the galaxy orbits, while the results shown in Fig.~\ref{integration_beta} are obtained out of all the galaxies under consideration. When computing the evolution of the anisotropy parameter for the bound subsample, the results for case B are practically identical (most of the orbits are bound, in this case); for case A, instead, the number of bound orbits is too small to derive significant prediction for the temporal evolution of $\beta$.
This little experiment seems to suggest that taking into account the evolution of the potential well of the host cluster, due to cosmological accretion, is enough to reproduce the trend in the anisotropy parameter as obtained from the MS for the sample of galaxies under consideration. This is possible notwithstanding the over-simplistic assumptions of a perfectly spherically-symmetric, smooth distribution and of an isolated, point-mass satellite. At the same time, we do not register a corresponding decrease in the mean circularity of the bound orbits, as one would naively expect, but rather an absence of significant evolution.
\section[]{Summary}
\label{sec:summary}
We have studied the evolution of the anisotropy parameter for galaxies orbiting within the most massive clusters extracted from the MS; we have related the value of this parameter to the internal properties of the galaxies, as predicted by the semi-analytic models of GUO11. Our findings can be summarised as follows:
\begin{itemize}
\item At redshift zero, blue galaxies move on less radial orbits than red galaxies do ($\beta_{blue} \approx 0.19$ vs. $\beta_{red} \approx 0.35$; see Fig.~\ref{beta_profile}).
\item At higher redshifts, progenitors of redshift-zero objects move on less radial orbits with respect to the full population of member galaxies (see Fig.~\ref{beta_highz_prog}); this is particularly true for progenitors of galaxies that are blue at redshift zero.
\item The orbits of infalling galaxies become increasingly radial going towards redshift zero (see Fig.~\ref{beta_infall_all}).
\item Of all the galaxies entering the cluster at a certain time, those that will survive until redshift zero are the subgroup characterised by the most tangential orbits at infall (see Fig.~\ref{beta_infall_prog}).
\item Of all the blue galaxies having a redshift-zero descendant (either red or blue) and entering the cluster at a certain time, those that remain blue until redshift zero are the subgroup of infalling objects on the most tangential orbits (compare the blue and cyan curves in Fig.~\ref{beta_infall_prog}).
\item The orbit of satellite galaxies become increasingly radial in time, especially just after the infall inside the cluster environment (see Fig.~\ref{beta_time}).
\item This increase can be reproduced by a simple integration of the galaxy orbit, taking into account the cosmological growth of the halo mass (see Sec.~\ref{sec:time_evol_beta}).
\end{itemize}
These findings and their statistical significance strongly suggest that the orbital features at infall have a major influence on the subsequent evolution of the galaxies, i.e. on their survival time within the cluster and on their turning from being star-forming/blue to passive/red objects. Galaxies on more markedly radial orbits will be more easily disrupted and removed from the cluster members as time goes by; even if they survive until redshift zero, these objects are those most likely to undergo changes in their internal properties and turn from being star-forming/blue to passive/red. Galaxies which, conversely, join their host with a significant component of their motion along the tangential direction tend to be less affected by the cluster environment; to this category belong all the blue galaxies found in the redshift-zero sample.\\

\section[]{Discussion}
\label{sec:discussion}
In the previous sections we have been drawing conclusions from a model which, albeit being state-of-the-art, is well known to contain some level of simplification in its treatment of galaxy formation and evolution; caution is therefore needed.\\
As discussed in Section 4.4 of \cite{guo2011}, only for galaxies in the stellar mass range $9.5 < \mathrm{log}(M_*/M_{\odot}) < 11$ does the model predict $u-i$ colour distribution in agreement with observations; at lower (higher) masses the galaxies tend to be too red (blue) and the colour distribution strongly deviates from the reference given by the SDSS/DR7 sample. These results were obtained when applying the model to both the MS and the Millennium-II Simulation \citep{mII}, selecting galaxies with $\mathrm{log}(M_*/M_{\odot}) > 10$ from the former and galaxies with $\mathrm{log}(M_*/M_{\odot}) \leq 10$ from the latter. We have performed our analysis on the full galaxy population and after applying a mass-cut leaving objects within the above ranges ($9.5 < \mathrm{log}(M_*/M_{\odot}) < 11$ and $10 < \mathrm{log}(M_*/M_{\odot}) < 11$) only; the results remain qualitatively the same. \\
As already mentioned, galaxies orphans of their dark-matter halo, or ``type $2$s'', were not included in the analysis due to the lack of information on their real dynamics. These objects are an important ingredient in the model as their presence allows the predicted galaxy luminosity function and radial number density profile in clusters to match observations; in addition to this, they represent a substantial fraction of the satellite populations (they account for half of all cluster members with $ M_* > 10^{10}M_{\odot}$). Including these objects in our sample has the immediate effect of increasing the median stellar age, while lowering the median specific star formation rate\footnote{The $u-i$ colour index increases, and correspondingly the sample becomes redder, at a given a stellar mass}; briefly, the sample ages. The effect on the anisotropy parameter corresponds to enhanced radial motions: from $\beta \approx 0.25$ for the full redshift-zero population without type $2$s, we move to $\beta \approx 0.35$ when these are included. The relative differences between the blue and red population are maintained. The reason for this substantial increase in radial anisotropy can be traced back to the origin of the orphan population itself; galaxies on more radial orbits are more subject to stripping and will lose their dark-matter component more easily. It may well be that a dark halo is still present in these objects, but that it consists of an insufficient number of particles to be regarded as a real sub-structure by the halo finders; in these cases, the trajectory of the most bound particle would still preserve the radial feature of the original satellite orbit. \\
We have split the original sample of galaxies into two different populations characterised by different values of the $u-i$ colour indicator. As a threshold, we have adopted the value where the colour distribution separates into two distinct component; for the galaxy sample we selected, this occurs at $u-i \approx 2.5$ at redshift zero. We have performed the same analysis changing this threshold and making more extreme cuts, both at redshift zero and at higher redshifts; the results remain coherent with those shown in the previous section. We have also used other properties, besides colour, to identify the two different galaxy populations, namely specific star formation rate and mean stellar age. Not surprisingly, as all these properties are expected to relate to one another, the results have not significantly changed.\\
We already mentioned that, according to our results, the type of orbit a galaxy is moving on when it last enters the cluster environment has major consequences on the evolution of the internal properties of the object. The finding must be related to the mechanisms responsible for these environment-induced changes affecting satellite galaxies moving within a cluster potential. The model of GUO11 features a more sophisticated and realistic treatment of these effects with respect to earlier attempts; indeed, as opposed to previous semi-analytic models \citep{baldry06, weinmann06_2, wang07, delucia07}, where the hot gas associated to a galaxy is immediately and entirely removed as the objects is accreted on to a larger system, in GUO11 the stripping of gas is performed gradually and modelled to reproduce the combined effect of tidal and ram-pressure forces (extending and integrating the recipes of \cite{font08} and \cite{weinmann10}). In addition to this, these processes are activated only while the satellite resides within the virial radius of the host; this limits the impact of possible failures in the FOF algorithm to identify physically independent, but spatially close, structures. The improved treatment of environmental effects allows GUO11 to reproduce, with noticeable accuracy, the radial distribution of star-forming, cluster galaxies as found in the SDSS data for a large sample of nearby clusters (see their Fig.~3).  Albeit its successes, the model is still incomplete as it lacks any treatment for the stripping of the cold-gas component, an effect known to be important in the inner regions of rich clusters \citep{gavazzi89, solanes01,boselli06}. Even though the overall treatment of environmental processes in the model does not entirely encompass the full range of mechanisms at work in real clusters, we think our results are not significantly affected; if anything, the inclusion of cold-gas stripping and therefore a more aggressive implementation of gas removal could only strengthen the differences found in the galaxy properties as a function of their initial orbit. \\
These speculations are confirmed by the results we found when applying a similar analysis to hydrodynamical simulations, where the co-evolution of dark and baryonic matter is followed self-consistently throughout time. We have considered few among the most massive and relaxed clusters from the Hutt\footnote{http://www.mpa-garching.mpg.de/$\scriptstyle\mathtt{\sim}$kdolag/Simulations/\label{fn:klaus}} sample \citep{dolag09}, a set of high-resolution, zoomed cluster simulation with implemented cooling, star-formation and feedback processes; we also used the results from the cosmological Magneticum Pathfinder Simulations\textsuperscript{\ref{fn:klaus}}, isolating the most massive clusters in the $128$ Mpc/h-side box (``Box $3$'' run) as well as in the $896$ Mpc/h-side box (``Box $1$a'' run). In all cases, we found anisotropy profiles in qualitative agreement with those from the semi-analytic models in terms of radial behaviour and global values for $\beta$. More importantly, we found that simulated galaxies characterised by either a young stellar population, low $u-i$ colour index or high specific star formation rate ($> 1 \times 10^{-11} M_{\odot}\; yr^{-1}$) present a systematically lower anisotropy parameter in each of the radial bins.\\
To summarise, we are confident that the intrinsic limits of the semi-analytic model by GUO11 do not affect the conclusions drawn in this paper.\\
A final comment on our choice regarding the cluster sample: we have focussed our analysis on the most massive clusters found in the MS - objects with virial masses greater than $2 \times 10^{14} \; M_{\odot}$ -  but we emphasise that there is no reason to expect the results obtained for this sample to apply at lower host masses. In fact, studying the properties of the dark-matter subhalos identified in the Millennium-II Simulation, \cite{faltenbacher10}  shows that their global anisotropy parameter depends both on the host mass and on the environment the host is sitting in, with satellites residing in either massive or isolated clusters moving on more markedly radial orbits. On the other hand, \cite{wetzel11} examines high-resolution, N-body simulation and evaluates the orbital parameters of satellites at the time of infall, defined as the first time the object crosses the virial radius of a larger host halo; again, a clear dependence of the initial orbit on the host mass is found, with galaxies characterised by increasingly radial motions at higher host halo masses. We therefore do not expect our results to apply to environments other than those of the most massive structures, at least not on a quantitative level; also the dependence of the internal evolution of the galaxy properties on the initial orbit may not be as strong in smaller objects or groups, where ram-pressure stripping is much less efficient a process than in rich clusters.\\
Besides the results on the link between orbital and internal properties of galaxies, we also showed our findings on the temporal evolution of the orbital anisotropy itself. We found that this quantity increases in time, especially just after infall. We note that our definition of infall time may have an influence in this analysis, at least on the magnitude of the infall anisotropy; half of the substructures hosting galaxies become part of the friend-of-friend halo of the cluster at distances exceeding two virial radii and may therefore have an extra tangential component in their initial velocity arising from the pull of the matter distribution at the outskirts of the host. We found that we can reproduce the increase in anisotropy by integrating the orbits of galaxies in a spherically-symmetric, smooth mass distribution, when the cosmological growth is taken into account. When investigating the evolution of the circularity distribution for the bound subsample of the integrated orbits, we found very little evolution. This suggests that the relation between the anisotropy parameter for a population of galaxies and the distribution of their circularities is less naive than a simple one-to-one correspondence.\\
We register some tension between our results and existing observations. Analysing two cluster sets at $z\approx 0$ and $z\approx 0.6$, \cite{BP09} depict a scenario where radially-infalling galaxies progressively turn from being star-forming to quiescent, while reducing their anisotropy from positive values down to $\beta \approx 0$. At redshift zero, the resulting anisotropy profile for the star-forming galaxy sample is consistent with more radially biased orbits than those characterising the quiescent counterpart. This is equivalent to saying that the orbits evolve in parallel with the internal properties of the satellites and that the former do not significantly impact the latter. Our analysis does, instead, support the opposite scenario, as thoroughly discussed. Earlier on, \cite{mahdavi99} and \cite{biviano04} reported of observational findings supporting more radial orbits for late-type galaxies than for early-type ones; their explanation was, again, in terms of coeval changes in the orbital and internal properties of the objects. Other observations, as mentioned in Sec.~\ref{sec:introduction}, report of gas-rich, late-type spirals in nearby clusters being found on more tangential orbits than early-type, gas-poor spirals \citep{dressler86,solanes01}; this is better reconciled with our findings, at least those regarding the impact of the satellites' orbits on their internal evolution. We note though, that these works agree in finding early-type galaxies (lenticulars and ellipticals) on close to ``isotropic'' orbits. \\
Both the observational and the numerical approach come with specific limitations and it would be interesting to pinpoint the origin of the disagreement. We leave this task to future works.

\section*{Acknowledgments}
The Millennium Simulation databases used in this paper and the web application providing online access to them were constructed as part
of the activities of the German Astrophysical Virtual Observatory. We
thank Gerard Lemson for his assistance in the use of the
database. We are grateful to the anonymous referee for his careful
reading of the manuscript. F.I. acknowledges useful discussions with
Andrea Biviano, Laura Sales and Chervin Laporte. K.D. acknowledges the support by the
DFG Priority Programme 1177 and additional support by the DFG Cluster
of Excellence ``Origin and Structure of the Universe''.

\bibliography{biblio}{}
\bibliographystyle{mn2e}

\label{lastpage}

\end{document}